\documentclass[journal,comsoc]{IEEEtran}
\usepackage[T1]{fontenc}% optional T1 font encoding
\usepackage{cite}
\usepackage{amsmath,amssymb,amsfonts}
\usepackage{algorithmic}
\usepackage[linesnumbered,ruled,vlined]{algorithm2e}
\usepackage{graphicx}
\usepackage{textcomp}
\usepackage{xcolor}
\usepackage{amsthm}
\theoremstyle{definition}
\newtheorem*{theorem*}{Theorem}
\usepackage{version}	% required for `\comment' (yatex added)
\begin{document}

\title{
Optimal Sequence and Performance for Desired User in Asynchronous CDMA System
}

\author{Hirofumi~Tsuda,~\IEEEmembership{Student Member,~IEEE,}
\thanks{This work was partially supported by JSPS (KAKENHI) Grant Number 18J12903. The material in this paper was presented in part at the EuCNC 2019, Valencia, Spain, June 2019.}
\thanks{The author is with the Department
of Applied Mathematics and Physics, Graduate School of Informatics, Kyoto University, Kyoto,
606-8501, Japan (e-mail: tsuda.hirofumi.38u@st.kyoto-u.ac.jp).}}

\maketitle

\begin{abstract}
We consider asynchronous CDMA systems in no-fading environments with a particular focus on a certain user. This certain user is called a desired user in this paper. In such a situation, an optimal sequence, maximum Signal-to-Interference plus Noise Ratio (SINR) and the maximum capacity for a desired user are derived with other spreading sequences being given and fixed. In addition, the maximum SINR and the optimal sequence for a desired user are written in terms of the minimum eigenvalue and the corresponding eigenvector of a matrix, respectively. Since it is not straightforward to obtain an explicit form of the maximum SINR, we evaluate SINR and obtain the lower and upper bounds of the maximum SINR. From these bounds, the maximum SINR may get larger as the quantities written in terms of quadratic forms of other spreading sequences decrease. Further, we propose a method to obtain spreading sequences for all the users which achieve large SINRs. The performance of our proposed method is numerically verified.
\end{abstract}


% creates the second title. It will be ignod for other modes.
\IEEEpeerreviewmaketitle

\section{Introduction}
\IEEEPARstart{T}{o} evaluate channel capacity is a significant task since channel capacity is the maximum achievable rate \cite{thomas_cover}. If the rate is smaller than given capacity, then there is a code whose maximum error converges to zero as the length of code words goes to infinity \cite{csiszar} \cite{gallager}. Thus, if large channel capacity is achieved, then information can be sent at a high rate. From such a reason, large capacity has been demanded. These results have been proven in \cite{shannon}. Furthermore, in a general channel, capacity has been obtained in \cite{general_channel} \cite{output_stat}. With a practical scheme, channel capacity has been evaluated in \cite{multilevel}. Further investigations of capacity are expected to contribute to improvement in communication systems. 

In Orthogonal Frequency Division Multiplexing (OFDM) systems, channel capacity with a non-linear amplifier has been obtained in \cite{clip}. In Multiple Input-Multiple Output (MIMO) systems, capacity has been investigated \cite{mimo}. By contrast, in some situations, capacity with MIMO systems has not been evaluated. 

In Code Division Multiple Access (CDMA) systems, capacity has also been evaluated. One of representative characteristics of CDMA systems is to use spreading sequences to communicate each other. Therefore, capacity may depend on spreading sequences. Further, it is known that capacity increases as Signal-to-Interference plus Noise Ratio (SINR) increases in practical schemes \cite{multilevel}. There are many works to obtain spreading sequences which achieve large SINR.

CDMA systems are divided into three kinds of systems. In synchronized CDMA systems, it is known that the Welch bound equality (WBE) sequences achieve the maximal capacity \cite{sync}. In chip synchronized CDMA systems with given and fixed delays, an algorithm to obtain sequences which achieves nearly maximum SINR has been suggested \cite{chipsync}. However, maximum channel capacity in asynchronous CDMA systems have not been evaluated. In asynchronous CDMA systems, many kinds of spreading sequences have been suggested to obtain large SINR. For more details, we refer the reader to \cite{mazzini}-\cite{gold} and asynchronous CDMA systems have been investigated in \cite{fundamental} \cite{criteria} \cite{pursley}. Since it is known that correlations play important roles in CDMA systems, correlations of sequences have been investigated and bounds of correlations have been obtained \cite{welch} \cite{sarwate}. Further, sequences which achieve the equalities of such bounds have been obtained \cite{zadoff} \cite{chu} \cite{meet}.  

In this paper, we show a optimal sequence for a desired user in a sense of SINR with no fading environments. Further, we show that the maximum SINR and an optimal sequence for a desired user are written in terms of the minimum eigenvalue and the corresponding eigenvector of a matrix, respectively. Since we derive the maximum SINR, the maximum capacity is derived under an approximation. Although we show an expression for the maximum SINR, it does not seem to be straightforward to obtain its closed form. To overcome this obstacle, we evaluate the maximum SINR and derive lower and upper bounds of SINR. From these bounds, it turns out that maximum SINR gets larger as the quantities written in terms of quadratic forms of other spreading sequences decrease. It is numerically verified that the maximum SINR for a desired user depends on the spreading sequences for the other users. From the derivation of the optimal sequence for a desired user, we propose a method to obtain spreading sequences for all the users which achieve large SINRs. In numerical results, we verify the performance of our method.

\section{System Description}
In this section, we show a model of asynchronous CDMA systems. This model has been investigated in \cite{pursley} \cite{borth} \cite{mypaper}. We make the following assumptions.
\begin{enumerate}
\item a modulation scheme is Binary Shift Phase Keying (BPSK)
\item there is no fading effect.
 \item the spreading sequences for the other users are given and fixed. Only a spreading sequence for a certain user is regarded as a variable. 
 \item channel noise follows Gaussian.
\item interference noise follows Gaussian.
\item interference noise is independent of Gaussian channel noise.
\item the phase of a transmitted signal, the time delay, and the transmitted symbols are random variables and uniformly distributed on their domains.
\end{enumerate}
The assumptions 1, 2 and 3 are often made and CDMA systems in no-fading effect have been investigated \cite{pursley} \cite{chipsync}. The assumptions 4 and 6 are usually made to analyze communication systems \cite{pursley} \cite{ofdmcdma}. The assumption 5 has been made in \cite{pursley}. Further, in analysis of Signal-to-Noise Ratio (SNR), this assumption is often made since Gaussian noise is the worst kind of additive noise in the view of capacity \cite{clip} \cite{gallager}. Thus, we consider the worst case in the view of capacity in asynchronous CDMA systems. The assumption 7 is often made to analyze asynchronous CDMA systems \cite{pursley} \cite{borth}.

Let $N$ be the length of spreading sequences and $N$ is common for all the users. From the assumption 1, a data signal of the user $k$, $b_k(t)$, is written as 
\begin{equation}
b_k(t) = \sum_{n=-\infty}^{\infty} b_{k,n} p_{T}(t - nT),
\end{equation}
where $b_{k,n} \in \{-1,1\}$ is the $n$-th component of the transmitted symbols which the user $k$ sends, $T$ is the duration of one symbol and $p_{T}(t)$ is the rectangular pulse written as
\[p_T(t) = \left\{ \begin{array}{c c}
1 & 0 \leq t < T\\
0 & \mbox{otherwise}
\end{array} \right. .\]
Then, the code waveform of the user $k$, $s_k(t)$, is written as
\begin{equation}
s_k(t) = \sum_{n=-\infty}^{\infty} s_{k,n} p_{T_c}(t - nT_c),
\end{equation}
where $s_{k,n}$ is the $n$-th component of the spreading sequence of the user $k$ and $T_c$ is the width of each chip such that $NT_c = T$. Here, we assume that the sequence $(s_{k,n})$ is periodic, that is, $s_{k,n}=s_{k,n+N}$. Moreover, we assume the power normalization condition
\begin{equation}
\sum_{n=1}^N \left|s_{k,n}\right|^2 = N.
\label{eq:power_cons}
\end{equation}
This condition is often used \cite{sarwate} \cite{welch}. With the above signals, the transmitted signal of the user $k$, $\zeta_k$, is written as
\begin{equation}
\zeta_k(t) = \sqrt{2P} \operatorname{Re}[s_k(t)b_k(t)\exp(j \omega_c t + j\theta_k)],
\label{eq:carrer}
\end{equation}
where $P$ is the common signal power to all the users, $\operatorname{Re}[z]$ is the real part of $z$, $j$ is the unit imaginary number, $\omega_c$ is the common carrier frequency to all the users and $\theta_k$ is the phase of the user $k$. Note that the signal $\zeta_k(t)$ is called a Radio Frequency (RF) signal.

We assume that there are $K$ users and that all the users are not synchronized. Then, the received signal $r(t)$ is written as
\begin{equation}
r(t) = \sum_{k=1}^K \zeta_k(t -\tau_k) + n(t),
\end{equation}
where $\tau_k$ is the delay time of the user $k$ and $n(t)$ is additive white Gaussian noise (AWGN). Note that the quantities $b_{k,n}$, $\theta_k$, and $\tau_k$ are random variables. 

To analyze SINR for a certain user, we focus on the user $i$ and the user $i$ is called desired user in this paper. If the user $i$ is a desired user and the received signal $r(t)$ is the input to a correlation receiver matched to $\zeta_i(t)$, then the corresponding output $Z_i$ is written as
\begin{equation}
Z_i = \int_{0}^T r(t) \operatorname{Re}[s_i(t-\tau_i)\exp(j \omega_c t + j \psi_i)]dt.
\label{eq:output}
\end{equation}
Without loss of generality, we assume $\tau_i = 0$ and $\theta_i = 0$. With a low-pass filter, we can ignore double frequency terms, and then rewrite Eq. (\ref{eq:output}) as
\begin{equation}
\begin{split}
Z_i &= \frac{1}{2} \sum_{k=1}^K \int_{0}^T \sqrt{2P} \operatorname{Re}[s_k(t)b_k(t)\overline{s_i(t)}\exp(j \psi_k)]dt\\
&+ \int_0^T n(t)\operatorname{Re}[s_i(t)\exp(j \omega_c t)]dt,
\label{eq:output2}
\end{split}
\end{equation}
where $\overline{z}$ is the complex conjugate of $z$,
\begin{equation}
\overline{s_i(t)} = \sum_{n=-\infty}^{\infty} \overline{s_{i,n}} p_{T_c}(t - nT_c),
\end{equation}
and $\psi_i = \theta_i - \omega_c\tau_i$.

In Eq. (\ref{eq:output2}), there are three kinds of random variables, the phases $\psi_k$, time delays $\tau_k$ and symbols $b_{k,n}$. From the assumption 7, these random variables, $\psi_k$, $\tau_k$ and $b_{k,n}$ are uniformly distributed on $[0, 2\pi)$, $[0,T)$ and $\{-1 ,1\}$, respectively. Without loss of generality, we assume that $b_{i,0} = +1$. To evaluate SINR, we define
\begin{equation}
\mu_{i,k}(\tau; t) = b_k(t - \tau)s_k(t - \tau)\overline{s_i(t)}.
\end{equation}
Then, the output value $Z_i$ is divided into three signals, the desired signal $D_i$, the interference signal $I_i$ and the AWGN signal $N_i$. They are written as
\begin{equation}
\begin{split}
D_i &= \sqrt{\frac{P}{2}} \int_0^T b_i(t)dt\\
I_i &= \sqrt{\frac{P}{2}} \sum_{\substack{k=1 \\ k \neq i}}\operatorname{Re}[\tilde{I}_{i,k}]\\
N_i &= \int_0^Tn(t)\operatorname{Re}[s_i(t)\exp(j\omega_ct)]
\end{split}
\end{equation}
where
\begin{equation*}
\tilde{I}_{i,k} = \int_0^T \mu_{i,k}(\tau_k; t) \exp(j \psi_k) dt.
\end{equation*}
Thus, the output $Z_i$ is rewritten as
\begin{equation}
Z_i = D_i + I_i + N_i.
\end{equation}
Note that the quantities $I_i$, and $N_i$ are random variables. Since $\operatorname{E}\{I_i\} = \operatorname{E}\{N_i\} = 0$ and $\displaystyle\operatorname{E}\{D_i\} = T\sqrt{P/2}$, we have $\displaystyle \operatorname{E}\{Z_i\} = T\sqrt{P/2}$, where $\operatorname{E}\{X\}$ is the mean of $X$. Then, SINR of the user $i$ is defined as
\begin{equation}
\operatorname{SINR}_i =\sqrt{\frac{PT^2/2}{\operatorname{Var}\{I_i\}  + \operatorname{Var}\{N_i\} }},
\label{eq:SINR_def}
\end{equation}
where $\operatorname{Var}\{X\}$ is the variance of $X$. From \cite{pursley} \cite{borth}, the variance of $N_i$ is written as
\begin{equation}
\operatorname{Var}\{N_i\} = \frac{1}{4}N_0T
\end{equation}
if $n(t)$ has a two-sided spectral density denoted as $\frac{1}{2}N_0$.

In \cite{mypaper}, the formula of SINR has been proposed as
\begin{equation}
 \operatorname{SINR}(\mathbf{s}_i)_i = \left\{ \frac{1}{6N^2}\sum_{\substack{k=1 \\ k \neq i}}^K \sum_{m=1}^N S_m^{i,k} + \frac{N_0}{2PT} \right\}^{-1/2},
\label{eq:SINR}
\end{equation}
where
\begin{equation}
 S_m^{i,k} =  \left( \mathbf{s}_i^* Q_m \mathbf{s}_i\right)\left( \mathbf{s}_k^* Q_m \mathbf{s}_k\right) + \left( \mathbf{s}_i^* \hat{Q}_m \mathbf{s}_i\right)\left(\mathbf{s}_k^* \hat{Q}_m \mathbf{s}_k \right).
\label{eq:quad_S}
\end{equation}
In this paper, attention is drawn to this formula. The symbols in Eq. (\ref{eq:SINR_def}) are explained as follows. First, $\mathbf{s}_k$ is the vector written as
\begin{equation}
 \mathbf{s}_k = (s_{k,1},s_{k,2},\ldots,s_{k,N})^\top,
\end{equation}
the matrices $Q_m$ and $\hat{Q}_m$ are given by
\begin{equation}
 Q_m = V^* C_m V, \hspace{3mm}\hat{Q}_m = \hat{V}^* \hat{C}_m \hat{V},
\label{eq:def_Q}
\end{equation}
where $V$ and $\hat{V}$ are unitary matrices whose $(m,n)$-th component is written as
\begin{equation}
\begin{split}
V_{m,n} &= \frac{1}{\sqrt{N}}\exp\left(-2 \pi j \frac{mn}{N}\right),\\
 \hat{V}_{m,n} &= \frac{1}{\sqrt{N}}\exp\left(-2 \pi j n\left(\frac{m}{N} + \frac{1}{2N}\right)\right),
\end{split}
\end{equation}
and $C_m$ and $\hat{C}_m$ are diagonal matrices whose $(m,m)$-th elements are given by
\begin{equation}
\begin{split}
\left(C_m\right)_{m,m} &= \sqrt{1+\frac{1}{2}\cos\left(2 \pi \frac{m}{N}\right)},\\
\left(\hat{C}_m\right)_{m,m} &= \sqrt{1+\frac{1}{2}\cos\left(2 \pi \left(\frac{m}{N} + \frac{1}{2N}\right)\right)},
\end{split}
\end{equation}
and the other elements are zero. In the above equations, $\mathbf{x}^\top$ and $\mathbf{z}^*$ denote the transpose of $\mathbf{x}$ and the conjugate transpose of $\mathbf{z}$, respectively. Note that the matrices $Q_m$ and $\hat{Q}_m$ are positive semidefinite matrices since $Q_m$ and $\hat{Q}_m$ are Gram matrices. It is obvious that Eq. (\ref{eq:SINR}) depends on the vector $\mathbf{s}_i$.

\section{Optimal Sequence and SINR for Desired User in No Fading}

In this section, we derive an optimal spreading sequence in no fading situation for the user $i$. Since the optimal spreading sequence is derived, the maximum SINR and the maximum capacity for the user $i$ are obtained.

In the previous section, we have made seven assumptions. These are also assumed in this section.
In this case where all the spreading sequence $\mathbf{s}_k$ are given, by assumption 1, 2, 3, and 6, SINR for the user $i$ is written as Eq. (\ref{eq:SINR}).
From assumption 3, Eq. (\ref{eq:SINR}) depends on only $\mathbf{s}_i$ since the other spreading sequences $\mathbf{s}_k$ are fixed for $k \neq i$. Therefore, to maximize SINR, we consider the following optimization problem
\begin{equation}
\begin{split}
(P_i)& \hspace{3mm} \min \hspace{2mm} \sum_{\substack{k=1 \\ k \neq i}}^K \sum_{m=1}^N S_m^{i,k} \\
& \mbox{subject to} \hspace{3mm} \|\mathbf{s}_i\|^2 = N,
\end{split}
\end{equation}
where $\|\mathbf{z}\|$ is the Euclidean norm of $\mathbf{z}$.
Note that the constraint is obtained from Eq. (\ref{eq:power_cons}).
It is clear that maximum SINR is obtained from the above optimization problem. In what follows, the problem $(P_i)$ is rewritten in another form. 

To analyze the optimization problem, we define the following matrix $\Sigma_i$
\begin{equation}
 \Sigma_i = \sum_{\substack{k=1 \\ k \neq i}}^K \sum_{m=1}^N \left( \mathbf{s}_k^* Q_m \mathbf{s}_k\right)Q_m + \left( \mathbf{s}_k^* \hat{Q}_m \mathbf{s}_k\right)\hat{Q}_m.
\label{eq:sigma}
\end{equation}
The matrix $\Sigma_i$ is constant since $\mathbf{s}_k$ is given and fixed for $k \neq i$ under assumption 3. Further, the matrix $\Sigma_i$ is positive semidefinite since the quantities $\left( \mathbf{s}_k^* Q_m \mathbf{s}_k\right)$ and $\left( \mathbf{s}_k^* \hat{Q}_m \mathbf{s}_k\right)$ are non-negative, and the matrices $Q_m$ and $\hat{Q}_m$ are positive semidefinite.

With the matrix $\Sigma_i$, the optimization problem $(P_i)$ is rewritten as
\begin{equation}
\begin{split}
(P_i)& \hspace{3mm} \min \hspace{2mm} \mathbf{s}_i^* \Sigma_i \mathbf{s}_i\\
& \mbox{subject to} \hspace{3mm} \|\mathbf{s}_i\|^2 = N.
\end{split}
\end{equation}
Further, the above problem is equivalent to the following one
\begin{equation}
\begin{split}
(P_i)& \hspace{3mm} \min \hspace{2mm} \frac{\mathbf{s}_i^* \Sigma_i \mathbf{s}_i}{\|\mathbf{s}_i\|^2/N}\\
& \mbox{subject to} \hspace{3mm} \|\mathbf{s}_i\|^2 = N.
\end{split}
\end{equation}
Let the vector $\mathbf{u}_i$ be $\mathbf{u}_i = \frac{1}{\sqrt{N}}\mathbf{s}_i$. With $\mathbf{u}_i$, the problem $(P_i)$ is rewritten as
\begin{equation}
\begin{split}
(P_i)& \hspace{3mm} \min \hspace{2mm} \frac{N \cdot \mathbf{u}_i^* \Sigma_i \mathbf{u}_i}{\|\mathbf{u}_i\|^2}\\
& \mbox{subject to} \hspace{3mm} \|\mathbf{u}_i\|^2 = 1.
\end{split}
\end{equation}
It is obvious that the value of the objective function is invariant under the action $\mathbf{u}_i \mapsto c \mathbf{u}_i$, where $c \in \mathbb{C}$ is a non-zero scalar. This observation yields that if we obtain a non-zero solution $\tilde{\mathbf{u}}'$ which minimizes the objective function of $(P_i)$, then we can obtain the feasible optimal solution $\tilde{\mathbf{u}}$ as $\tilde{\mathbf{u}} = \tilde{\mathbf{u}}'/\|\tilde{\mathbf{u}}'\|$.
Thus, we consider the following problem
\begin{equation}
\begin{split}
(P'_i)& \hspace{3mm} \min_{\mathbf{u}_i \neq \mathbf{0}} \hspace{2mm} \frac{N \cdot \mathbf{u}_i^* \Sigma_i \mathbf{u}_i}{\|\mathbf{u}_i\|^2}.
\end{split}
\end{equation}
This is the Rayleigh quotient of $N\Sigma_i$ \cite{manifold}. It is known that the optimal value coincides with the product of $N$ and the minimum eigenvalue of $\Sigma_i$, $\lambda^{(i)}_{\min} \geq 0$, and that the global minimizer of the problem $(P_i)$ is the eigenvector corresponding to $\lambda^{(i)}_{\min}$. Let $\mathbf{u}$ be such a minimizer. When the minimizer $\mathbf{u}_i$ is normalized as $\|\mathbf{u}_i\|=1$, the optimal spreading sequence for the user $i$, $\mathbf{s}_i^\star$, is written as
\begin{equation}
 \mathbf{s}_i^\star = \sqrt{N}\mathbf{u}_i.
\end{equation}
Then, the maximum SINR is written as
\begin{equation}
 \operatorname{SINR}_i^\star = \operatorname{SINR}(\mathbf{s}_i^\star)_i = \left\{\frac{\lambda^{(i)}_{\min}}{6N} +\frac{N_0}{2PT}\right\}^{-1/2}.
\label{eq:optimal_SINR_i}
\end{equation}
Further, it is known that the channel capacity is written in terms of Signal-to-Noise Ratio (SNR) if an input is continuous and channel noise is Gaussian \cite{thomas_cover} \cite{el_gamal}. The sum of interference noise and channel noise follow Gaussian since the sum of the two independent Gaussian variables follow Gaussian under assumptions 5 and 6 \cite{prob}. Even in a case where noise follows Gaussian, the channel capacity with a practical scheme is complicated \cite{gallager} \cite{multilevel}. In \cite{multilevel}, the channel capacity with BPSK scheme is close to one with a continuous channel in low SNR. Taking into account these reasons, we approximate the maximum channel capacity of the user $i$ by one with a continuous channel. Under this approximation, from Eq. (\ref{eq:optimal_SINR_i}), the maximum channel capacity for the user $i$, $C^\star_i$, is evaluated as
\begin{equation}
 C^\star_i \approx \frac{1}{2}\log\left[1 +  \left\{\frac{\lambda^{(i)}_{\min}}{6N} +\frac{N_0}{2PT}\right\}^{-1}\right].
\label{eq:opt_capacity}
\end{equation}
As seen in  the above discussions, the maximum SINR and the maximum channel capacity depend on the minimum eigenvalue of the matrix $\Sigma_i$, and these maximums are achieved with the eigenvector corresponding to the minimum eigenvalue.

\section{Estimating Maximum SINR}
We have evaluated the maximum SINR in asynchronous CDMA systems for a desired user. Since the matrix $\Sigma_i$ depends on $\mathbf{s}_k$ $(k \neq i)$, the minimum eigenvalue $\lambda^{(i)}_{\min}$ may depend on other spreading sequences $\mathbf{s}_k$ $(k \neq i)$. Thus, to analyze the maximum SINR, it is necessary to obtain the explicit form of $\lambda^{(i)}_{\min}$. However, it is not straightforward to obtain the explicit form of $\lambda^{(i)}_{\min}$. Instead, we derive the lower and upper bounds of the maximum SINR in this section. From these bounds, we can estimate the maximum SINR and the know what the dominant factor related to SINR is. 

As seen in Eq. (\ref{eq:sigma}), the matrix $\Sigma_i$ consists of two kinds of the matrices, $Q_m$ and $\hat{Q}_m$. From Eq. (\ref{eq:def_Q}), the eigenvalues of the matrices $Q_m$ and $\hat{Q}_m$ are represented as the matrices $C_m$ and $\hat{C}_m$, respectively. Further, the matrices $C_m$ and $\hat{C}_m$ have one non-zero component at the $(m,m)$-th entry. Therefore, the matrix $\Sigma_i$ is written as
\begin{equation}
 \Sigma_i = V^* \Lambda_i V + \hat{V}^* \hat{\Lambda}_i \hat{V},
\label{eq:decomp_sigma}
\end{equation} 
where $\Lambda_i$ and $\hat{\Lambda}_i$ are diagonal matrices whose $m$-th diagonal components, $\lambda^{(i)}_m$ and $\hat{\lambda}^{(i)}_m$, are written as
\begin{equation}
\begin{split}
 \lambda^{(i)}_m &= \sqrt{1+\frac{1}{2}\cos\left(2 \pi \frac{m}{N}\right)} \sum_{\substack{k=1 \\ k \neq i}}^K \left( \mathbf{s}_k^* Q_m \mathbf{s}_k\right)\\
\hat{\lambda}^{(i)}_m &=  \sqrt{1+\frac{1}{2}\cos\left(2 \pi \left(\frac{m}{N} + \frac{1}{2N}\right)\right)}  \sum_{\substack{k=1 \\ k \neq i}}^K \left( \mathbf{s}_k^* \hat{Q}_m \mathbf{s}_k\right).
\end{split}
\end{equation}
Since the matrices $V$ and $\hat{V}$ are unitary, the quantities $\lambda^{(i)}_m$ and $\hat{\lambda}^{(i)}_m$ are the eigenvalues of the matrices $V^*\Lambda_i V$ and $\hat{V}^*\hat{\Lambda}_i\hat{V}$, respectively. Note that the quantities $\lambda^{(i)}_m$ and $\hat{\lambda}^{(i)}_m$ depend on the spreading sequences $\mathbf{s}_k$ for $k \neq i$.

With the above eigenvalues, the bounds of the maximum SINR for the user $i$ are derived. First, we derive the upper bound. As seen in Eq. (\ref{eq:optimal_SINR_i}), the maximum SINR is written with the minimum eigenvalue of $\Sigma_i$, $\lambda^{(i)}_{\min}$. Since $\lambda^{(i)}_{\min}$ is the optimal value of the Rayleigh quotient of $\Sigma_i$, the following relations are obtained
\begin{equation}
 \begin{split}
  \lambda^{(i)}_{\min} =& \min_{\mathbf{u} \neq \mathbf{0}} \frac{\mathbf{u}^*\Sigma_i \mathbf{u}}{\|\mathbf{u}\|^2}\\
=& \min_{\mathbf{u} \neq \mathbf{0}} \frac{\mathbf{u}^* \left(V^* \Lambda_i V + \hat{V}^* \hat{\Lambda}_i^* \hat{V}\right) \mathbf{u}}{\|\mathbf{u}\|^2}\\
=& \min_{\substack{\mathbf{u}_1 \neq \mathbf{0}, \mathbf{u}_2 \neq \mathbf{0} \\ \mathbf{u}_1 = \mathbf{u}_2}}\left[ \frac{\mathbf{u}^*_1 V^* \Lambda_i V \mathbf{u}_1}{\|\mathbf{u}_1\|^2} +  \frac{\mathbf{u}_2^* \hat{V}^* \hat{\Lambda}^*_i \hat{V} \mathbf{u}_2}{\|\mathbf{u}_2\|^2}\right]\\
\geq&  \min_{\mathbf{u}_1 \neq \mathbf{0}} \frac{\mathbf{u}^*_1 V^* \Lambda_i V \mathbf{u}_1}{\|\mathbf{u}_1\|^2} +  \min_{\mathbf{u}_2 \neq \mathbf{0}} \frac{\mathbf{u}_2^* \hat{V}^* \hat{\Lambda}^*_i \hat{V} \mathbf{u}_2}{\|\mathbf{u}_2\|^2}\\
=& \min_m \lambda^{(i)}_m + \min_m \hat{\lambda}^{(i)}_m,
 \end{split}
\label{eq:lower_lambda}
\end{equation}
where we have used Eq. (\ref{eq:decomp_sigma}) and the inequality in Eq. (\ref{eq:lower_lambda}) is established since the feasible region gets larger. Then, the upper bound of the maximum SINR is written as
\begin{equation}
 \left\{\frac{1}{6N}(\min_m \lambda^{(i)}_m + \min_m \hat{\lambda}^{(i)}_m) + \frac{N_0}{2PT}\right\}^{-1/2} \geq \operatorname{SINR}^\star_i.
\end{equation}
On the other hand, to derive the lower bound of the maximum SINR, we use the following theorem \cite{Weyl_eigen}.
\begin{theorem*}[{\bf Weyl}]
Let $A$ and $B$ be the $n \times n$ Hermitian matrices whose eigenvalues are written as $\alpha_1 \geq \alpha_2 \geq \cdots \geq \alpha_n$ and $\beta_1 \geq \beta_2 \geq \cdots \geq \beta_n$, respectively. Further, we define the Hermitian matrix $C=A+B$ whose eigenvalues are written as $\gamma_1 \geq \gamma_2 \geq \cdots \geq \gamma_n$. Then, the following relation holds for $k + l -1 \leq n$
\begin{equation}
 \gamma_{k+l-1} \leq \alpha_k + \beta_l.
\end{equation}
\end{theorem*}
The eigenvalues of sums of Hermitian matrices have been investigated in \cite{wielandt} \cite{fulton} \cite{horn}.
From the above theorem, it follows that
\begin{equation}
 \lambda^{(i)}_{\min} \leq \min \left\{\min_m \lambda^{(i)}_m+ \max_m \hat{\lambda}^{(i)}_m, \max_m \lambda^{(i)}_m + \min_m \hat{\lambda}^{(i)}_m\right\}.
\label{eq:upper_lambda}
\end{equation}
Equation (\ref{eq:upper_lambda}) is obtained when we set $(k,l)=(n,1)$ and $(k,l)=(1,n)$ in the theorem. With Eq. (\ref{eq:upper_lambda}), a lower bound of the maximum SINR is written as
\begin{equation}
 \left\{\frac{1}{6N} \gamma + \frac{N_0}{2PT}\right\}^{-1/2} \leq \operatorname{SINR}^\star_i,
\end{equation}
where
\begin{equation}
 \gamma =  \min \left\{\min_m \lambda^{(i)}_m+ \max_m \hat{\lambda}^{(i)}_m, \max_m \lambda^{(i)}_m + \min_m \hat{\lambda}^{(i)}_m\right\}.
\end{equation}
From Eqs. (\ref{eq:lower_lambda}) and (\ref{eq:upper_lambda}), we observe that the maximum SINR is related to the quantities $\lambda_m$ and $\hat{\lambda}_m$, that is, the maximum SINR for a desired user may depend on the sequences for the other users. This relation is numerically verified in Section VI. These observations yield that the maximum SINR is improved if the quantities $\lambda^{(i)}_m$ and $\hat{\lambda}^{(i)}_m$ are reduced. Therefore, if the spreading sequences $\mathbf{s}_k$ for $k \neq i$ are designed to achieve lower $\lambda^{(i)}_m$ and $\hat{\lambda}^{(i)}_m$ for $m=1,2,\ldots,N$, then larger SINR is obtained with the optimal sequence for the user $i$, $\mathbf{s}^\star_i$. 

\section{Algorithm to Obtain Large SINRs}
In the previous sections, we have discussed the SINR and capacity for a certain user with the optimal sequence. In this section, we discuss the way to obtain sequences for all the users which achieve large SINRs. To take into account sequences for all the users, we consider the following sum of the squared SINRs
\begin{equation}
\frac{1}{K}\sum_{i=1}^K \left( \operatorname{SINR}(\mathbf{s}_i)_i \right)^2 = \frac{1}{K}\sum_{i=1}^K \left\{  \frac{1}{6N^2}\sum_{\substack{k=1 \\ k \neq i}}^K \sum_{m=1}^N S_m^{i,k} + \frac{N_0}{2PT} \right\}^{-1}.
\label{eq:sum_squared_SINR}
\end{equation}
Note that the above quantity is the average of the squared SINRs, and is expected to yield SINR for all the users. Then, our goal is to obtain the sequences which make the quantity shown in Eq. (\ref{eq:sum_squared_SINR}) large.

However, it is not straightforward to analyze Eq. (\ref{eq:sum_squared_SINR}) since there is the sum of inverse numbers. To overcome this obstacle, we consider the harmonic mean of squared SINRs which is written as
\begin{equation}
\begin{split}
&K \left\{ \sum_{i=1}^K \left( \operatorname{SINR}(\mathbf{s}_i)_i \right)^{-2} \right\}^{-1} \\
=& K \left\{  \frac{1}{6N^2} \sum_{i=1}^K \sum_{\substack{k=1 \\ k \neq i}}^K \sum_{m=1}^N S_m^{i,k} + \frac{KN_0}{2PT} \right\}^{-1}.
\label{eq:harm_ave_SINR}
\end{split}
\end{equation}
From the relation between the arithmetic mean and the harmonic mean, the following relation is established
\begin{equation}
\begin{split}
& K \left\{  \frac{1}{6N^2} \sum_{i=1}^K \sum_{\substack{k=1 \\ k \neq i}}^K \sum_{m=1}^N S_m^{i,k} + \frac{KN_0}{2PT} \right\}^{-1}\\
\leq & \frac{1}{K}\sum_{i=1}^K \left\{  \frac{1}{6N^2}\sum_{\substack{k=1 \\ k \neq i}}^K \sum_{m=1}^N S_m^{i,k} + \frac{N_0}{2PT} \right\}^{-1}.
\label{eq:harm_ave_SINR}
\end{split}
\end{equation}

From the above inequality, it is expected that the average of the SINRs increases as the harmonic mean increases. Thus, instead of the average of SINRs, we consider the harmonic mean of SINRs. Then, we consider the following problem
\begin{equation}
\begin{split}
(P)& \hspace{3mm} \min \hspace{2mm} \sum_{i=1}^K \sum_{\substack{k=1 \\ k \neq i}}^K \sum_{m=1}^N S_m^{i,k} \\
& \mbox{subject to} \hspace{3mm} \|\mathbf{s}_i\|^2 = N\hspace{3mm}(i=1,\ldots,K).
\end{split}
\end{equation}
Similar to the discussion in Section III, we consider only the user $i$. Here, we assume that only the sequence for the user $i$, $\mathbf{s}_i$, is a variable and that the other sequences $\mathbf{s}_k$ are given and fixed for $k \neq i$. This idea is seen as an alternating direction method of multipliers (ADMM) technique \cite{admm}. Under this assumption, we solve the following problem
\begin{equation}
\begin{split}
(P_i)& \hspace{3mm} \min \hspace{2mm} \sum_{\substack{k=1 \\ k \neq i}}^K \sum_{m=1}^N S_m^{i,k} \\
& \mbox{subject to} \hspace{3mm} \|\mathbf{s}_i\|^2 = N.
\end{split}
\end{equation}
We emphasize that the other sequences $\mathbf{s}_k$ for $k \neq i$ are given and fixed (see assumption 3 in Section II). As seen in Section III, the optimal value and minimizer are written in terms of the minimum eigenvalue and the corresponding eigenvector of the matrix $\Sigma_i$, respectively. Our algorithm is written in Algorithm \ref{algo:ours}.
\begin{algorithm}[h]
Set the initial sequences $\mathbf{s}_k$ for $k=1,\ldots,K$ and $l=0$. Set $L \geq 1$.\\
For $k=1,\ldots,K$, solve the problem $(P_i)$, obtain the optimal solution $\mathbf{s}^\star_k$, and set $\mathbf{s}_k \leftarrow \mathbf{s}^\star_k$.\\
$l \leftarrow l+1$.\\
If $\{ \mathbf{s}_k \}_{k=1,\ldots,K}$ converge or $l = L$, then go to Step 5. Otherwise, go to step 2.\\
Output $\mathbf{s}_k$.
\caption{Algorithm to obtain sequences which achieve large SINRs.}
\label{algo:ours}
\end{algorithm}
Note that when the problem $(P_i)$ is solved, the sequences $\mathbf{s}_k$ ($k=1,\ldots,i-1$) have already been updated.

Here we give an explanation about why large SINRs will be achieved with Algorithm \ref{algo:ours}. As seen in the problem $(P)$, our aim is to achieve the large harmonic mean of squared SINRs. For $i$, the objective function of the problem $(P)$ is evaluated as 
\begin{equation}
\begin{split}
\sum_{i=1}^K \sum_{\substack{k=1 \\ k \neq i}}^K \sum_{m=1}^N S_m^{i,k} &=  2\sum_{\substack{k=1 \\ k \neq i}}^K \sum_{m=1}^N S_m^{i,k} + \sum_{\substack{k_1=1 \\ k_1 \neq i}}^K \sum_{\substack{k_2=1 \\ k_2 \neq i,k_1}}^K \sum_{m=1}^N S_m^{k_1,k_2}\\
& = 2 \mathbf{s}^*_i \Sigma_i \mathbf{s}_i +  \sum_{\substack{k_1=1 \\ k_1 \neq i}}^K \sum_{\substack{k_2=1 \\ k_2 \neq i,k_1}}^K \sum_{m=1}^N S_m^{k_1,k_2}\\
& \geq  2 N \lambda^{(i)}_{\min} +  \sum_{\substack{k_1=1 \\ k_1 \neq i}}^K \sum_{\substack{k_2=1 \\ k_2 \neq i,k_1}}^K \sum_{m=1}^N S_m^{k_1,k_2},
\end{split}
\label{eq:decomp_harm}
\end{equation}
where we have used the fact that $S_m^{i,k} = S_m^{k,i}$ for all $m$ and the results obtained in Section III. In the right hand side of the first line in Eq. (\ref{eq:decomp_harm}), the first term depends on $\mathbf{s}_i$ and the last term is independent of $\mathbf{s}_i$. Further, the first term is the objective function of the problem $(P_i)$. Thus, the first term can be minimized with the sequence $\mathbf{s}^\star_i$ and its value equals to $2N\lambda^{(i)}_{\min}$. This observation yields that solving the problem $(P_i)$ is equivalent to minimizing the terms relating the sequence $\mathbf{s}_i$ in the harmonic mean of squared SINRs.

From the above discussions, it has been shown that solving the problem $(P_i)$ leads to reducing the harmonic mean of squared SINRs.

%%%%%%%%%%%%%%%%%%%%%%%%%%%%%%%%%%%%%%%%%%%%%%%%%%%%%%%%%%%%

\section{Numerical Results}
We obtain the sequences $\mathbf{s}_k$ for $k=1,\ldots,K$ with Algorithm \ref{algo:ours}. We set the number of users $K = 7$ and the length of sequences $N=31$. As the initial sequences (Step 1 in Algorithm \ref{algo:ours}), the Gold codes \cite{gold} and random sequences are used. We calculate Bit Error Rate (BER) as
\begin{equation}
 \mbox{BER} = \frac{1}{K}\frac{1}{U} \sum_{k=1}^K \sum_{u=1}^U \mbox{BER}_{k,u},
\end{equation}
where $\mbox{BER}_{k,u}$ is the BER of the user $k$ at the $u$-th iteration and $U$ is the number of iterations. Here, we set $U=1.0 \times 10^4$. As seen in Section II, the modulation scheme is BPSK. There is no fading effect.

\begin{figure}[htbp]
\centering  
\includegraphics[width=2.8in]{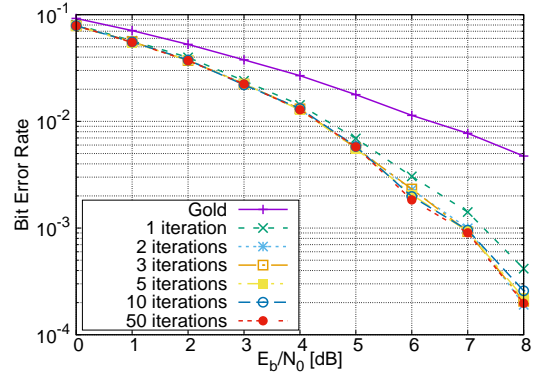}
\caption{Bit Error Rate with sequences of each iteration: Initial sequences are the Gold codes.}
\label{fig:ber_gold}
\end{figure}

Figure \ref{fig:ber_gold} shows the BER in the case where gold codes are used as the initial sequences. Here, $E_b$ denotes the average power per bit. Further, in the legend, ``iteration'' means $L$ in Algorithm \ref{algo:ours}. As seen in Fig. \ref{fig:ber_gold}, BER gets reduced when the number of iterations gets large. This observation yields that reducing the harmonic mean of squared SINRs leads to enlarging SINR for each user. In particular, the BER with one iteration is larger than ones with the other iterations. This observation yields that the maximum SINR for a desired user depends on the sequences for the other users. The reason is as follows. As seen in Eqs. (\ref{eq:lower_lambda}) and (\ref{eq:upper_lambda}), the maximum SINR for a desired user is written in terms of the sequences for the other users. In step 2 in our algorithm, the optimal sequence for the user $k$ is obtained. Then, SINR for the user $k$ is maximized. If the maximum SINR for a desired user is independent of the sequences for the other users, then the maximum SINR is constant for every iteration number $L$. However, as seen in Fig. \ref{fig:ber_gold}, it is observed that the BER is varied for every iteration number $L$. This yields that the maximum SINR is varied for every $L$. From the above discussions, it is numerically verified that the maximum SINR for a desired user depends on the sequences for the other users.

\begin{figure}[htbp]
\centering  
\includegraphics[width=2.8in]{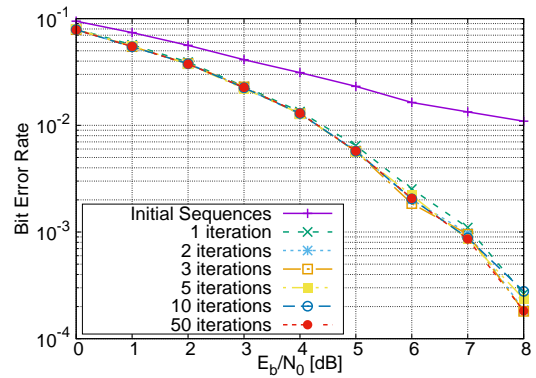}
\caption{Bit Error Rate with sequences of each iteration: Initial sequences are generated randomly.}
\label{fig:ber_random}
\end{figure}

Figure \ref{fig:ber_random} show the BER in the case where random sequences are used as the initial sequences. The aim is to verify whether the performance depends on initial sequences or not. As seen in Fig. \ref{fig:ber_random}, BER gets smaller as the number of iterations increases. Since the initial sequences are generated randomly, the initial sequences have large BER. However, the BER with one iteration reduces significantly. The BER with 50 iterations is the smallest in this figure and its value is nearly equivalent to one with 50 iterations in Fig. \ref{fig:ber_gold}. From this observation, it is expected that our algorithm can always achieve low BER when the number of iterations is sufficiently large. Note that we have not proven the convergence of the objective function of the problem $(P)$.

\begin{figure}[htbp]
\centering  
\includegraphics[width=2.8in]{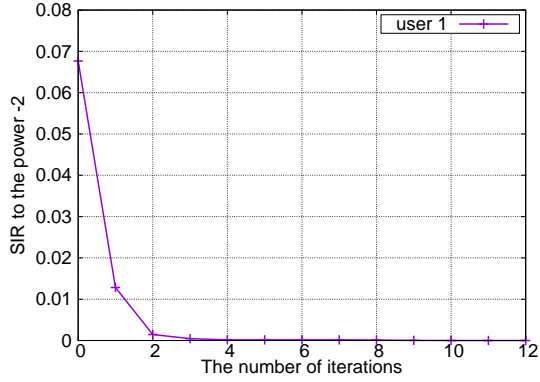}
\caption{SIR for the user 1 at each iteration: Initial sequences are the Gold codes.}
\label{fig:sir_0}
\end{figure}

\begin{figure}[htbp]
\centering  
\includegraphics[width=2.8in]{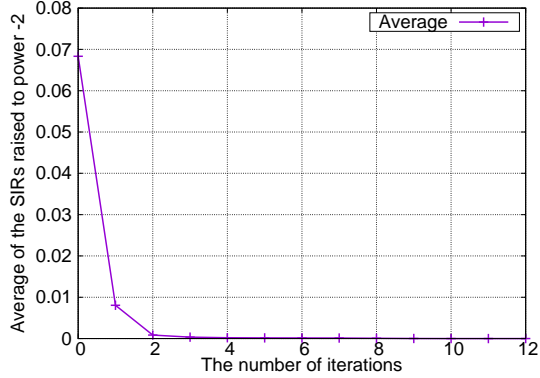}
\caption{Average of inverses of squared SIRs: Initial sequences are the Gold codes.}
\label{fig:sir_ave}
\end{figure}

Figures \ref{fig:sir_0} and \ref{fig:sir_ave} show the Signal-to-Interference noise Ratio (SIR) obtained with our algorithm. Here, SIR of the user $i$ is defined as
\begin{equation} 
\operatorname{SIR}_i= \operatorname{SIR}(\mathbf{s}_i)_i = \left\{ \frac{1}{6N^2}\sum_{\substack{k=1 \\ k \neq i}}^K \sum_{m=1}^N S_m^{i,k}\right\}^{-1/2}.
\label{eq:SIR_i} 
\end{equation} 
By definition, SIR is equivalent to SINR with $N_0 = 0$. To depict these two figures, the Gold codes are used as the initial sequences. Figure \ref{fig:sir_0} shows the SIR raised to the power -2 for the user 1 at each iteration in our algorithm, that is, the vertical axis shows $\operatorname{SIR}^{-2}_1$. Since this quantity is in the objective function of the problem $(P_1)$, this figure also shows the value of the objective function of the problem $(P_1)$. As seen in this figure, the inverse of the squared SIR at 1 iteration is larger than the others except for the original one (SIR with original sequences). From this observation, SIR of the optimal sequence for a desired user depends on the sequences for the other users. This result explains why the BER at 1 iteration is larger than the other ones except for the Gold codes (see Fig. \ref{fig:ber_gold}). Further, as the number of the iteration gets larger, the quantity $\operatorname{SIR}^{-2}_1$ gets closer to 0.

Figure \ref{fig:sir_ave} shows the average of the inverses of squared SIRs at each iteration, that is, the vertical axis in Fig. \ref{fig:sir_ave} shows
\begin{equation}
 \begin{split}
  &\mbox{(vertical axis in Fig. \ref{fig:sir_ave})}\\
=&\frac{1}{K}\sum_{i=1}^K\operatorname{SIR}^{-2}_i = \frac{1}{K}\sum_{i=1}^K  \frac{1}{6N^2}\sum_{\substack{k=1 \\ k \neq i}}^K \sum_{m=1}^N S_m^{i,k}.
 \end{split}
\label{eq:vert_fig2}
\end{equation}
This quantity is in the objective function of the problem $(P)$. As seen in  Fig. \ref{fig:sir_ave}, the value of Eq. (\ref{eq:vert_fig2}) gets smaller and closer to 0 as the number of iterations gets larger. In Fig. \ref{fig:sir_0}, we have seen that $\operatorname{SIR}^{-2}_1$ gets closer to 0 as the number of iterations gets larger. In Section V, we have considered the problem $(P)$ to take into account the SINRs for all the users. Further, from the relation between the arithmetic mean and the harmonic mean (see Eq. (\ref{eq:harm_ave_SINR})), when the quantity shown in Fig. \ref{fig:sir_ave} and Eq. (\ref{eq:vert_fig2}) gets reduced, the arithmetic mean of squared SIRs gets large. Thus, Fig. \ref{fig:sir_ave} numerically verifies that our algorithm can achieve large SIR for each user $i$ and large SINRs for all the users are achieved with our algorithm.

\section{Conclusion}
In this paper, we have derived the optimal spreading sequence for the user $i$, which achieves maximum SINR and maximum capacity under an approximation. It has turned out that the maximum SINR is written in terms of the minimum eigenvalue of the matrix $\Sigma_i$ and that the optimal spreading sequence is obtained as a corresponding eigenvector. Further, we have derived the lower and upper bounds of maximum SINR. From these bounds, the maximum SINR will get larger as the quantities $\lambda^{(i)}_m$ and $\hat{\lambda}^{(i)}_m$ get smaller. From the derivation of the optimal sequence for a desired user, we have proposed the algorithm to obtain the sequences which achieve large SINRs for all the users. In numerical results, the performance of our algorithm has been verified. These results have also shown that the performance of the optimal sequence for a desired user depends on the sequences for the other users.

To consider the practical situations, we have to take into account fading effects. One issue is to derive optimal sequences in a sense of SINR under fading effects. This should be considered somewhere as a remaining issue.

\section*{Acknowledgment}
The author would like to thank Dr. Shin-itiro Goto for his advise.

%%%%%%%%%%%%%%%%%%%%%%%%%%%%%%%%%%%%%%%%%%%%%%%5
 
%%%%%%%%%%%%%%%%%%%%%%%%%%%%%%%%%%%%%%%%

\end{document}